\documentstyle[11pt,newpasp,twoside,epsf]{article}
\reversemarginpar

\begin{document}
\title{The Starlab Environment for Dense Stellar Systems}
\author{Piet Hut}
\affil{Institute for Advanced Study, Princeton, NJ 08540, U.S.A.}

\begin{abstract}
Traditionally, a simulation of a dense stellar system required
choosing an initial model, running an integrator, and analyzing the
output.  Almost all of the effort went into writing a clever
integrator that could handle binaries, triples and encounters between
various multiple systems efficiently.  Recently, the scope and
complexity of these simulations has increased dramatically, for three
reasons: 1) the sheer size of the data sets, measured in Terabytes,
make traditional `awking and grepping' of a single output file
impractical; 2) the addition of stellar evolution data brings
qualitatively new challenges to the data reduction; 3) increased
realism of the simulations invites realistic forms of `SOS':
Simulations of Observations of Simulations, to be compared directly
with observations.  We are now witnessing a shift toward the
construction of archives as well as tailored forms of visualization
including the use of virtual reality simulators and planetarium domes,
and a coupling of both with budding efforts in constructing virtual
observatories.  This review describes these new trends, presenting
Starlab as the first example of a full software environment for
realistic large-scale simulations of dense stellar systems.

\end{abstract}

\section{A Brief History of Simulations of Dense Stellar Systems}

Modeling a star cluster on a computer is a subject with a published
history of more than forty years, starting with the 10-body
simulations by von Hoerner (1960).  These pioneering calculations had
to be halted when the first binary was formed.  In those days neither
was the hardware fast enough, nor the software sophisticated enough to
follow the long-term evolution of a perturbed binary.  Rapid
improvement of both hardware and software allowed the treatment of
dynamically formed binaries by the mid-sixties, but the maximum number
of bodies remained in the regime $N<100$ where relaxation effects
could not be ignored on a crossing time.  By the early seventies,
larger systems could be modeled, up to $N=500$.  By that time a clear
separation between relaxation time and crossing time could be
observed.

These simulations showed that the core of a typical star cluster
contracts as a result of heat flowing toward its outer regions, on a
two-body relaxation time scale.  This phenomenon had been predicted
theoretically and was dubbed `gravothermal collapse' or `core
collapse', where the term `collapse' is perhaps a bit misleading since
the process takes billions of years for a typical globular cluster.
It was observed in the simulations that this type of collapse
continues until a hard binary appears in the center, where the term
`hard' indicates that the binary has a binding energy much larger than
the typical kinetic energy of a single star.  Such binaries tend to
give off energy to passing stars, hardening even further while heating
their surroundings.  In this way, one or more hard binaries were
observed to provide the heat source that balanced the heat losses
further out in the star cluster, thus ending core contraction.  Just
as an individual star like the sun balances radiation losses from its
photosphere with nuclear energy generation in its core, after the
initial contraction of the protostar toward the main sequence, star
clusters with a `collapsed core' can similarly balance their energy
budget.

Key ingredients in making it possible to integrate these larger
500-body systems in the seventies were the use of individual time
steps (Aarseth 1963), as well as special treatments of binaries
through various ways of analytical and other forms of regularization
({\it cf.}\ Aarseth 1985, 2002).  By the late eighties the mechanism
of core collapse had been fully understood, including the conditions
for the occurrence of post-collapse core oscillations, using numerical
simulations in the conducting gas sphere approximation (Sugimoto \&
Bettwieser 1983, Bettwieser \& Sugimoto 1984), Fokker-Planck
approximation (Cohn {\it et al.} 1989, Murphy {\it et al.} 1990) as
well as semi-analytical arguments (Goodman 1987).  The first
demonstration of the occurrence of core oscillations in direct
$N$-body systems was given by Makino (1996), using the GRAPE-4 to
perform a 32,000-body calculation, a {\it tour de force} around that
time.  Only now, six years later, are simulations of this size
becoming more routine.  For a review of the GRAPE project, see Makino
(2002) and Makino \& Taiji (1998).

From an astrophysics point of view, $N$-body simulations of star
clusters have finally started to become realistic, over the last few
years, now that stellar evolution recipes have been tightly coupled to
stellar dynamical calculations.  The talks at this symposium by Hurley
(2002) and McMillan (2002) give pointers to the literature.  For more
general background information, see Heggie \& Hut (2002).  Most
interesting and most challenging are the treatments of multiple
stellar encounters: interactions between binaries and single stars,
binary-binary encounters, or even more complex interactions involving,
say, an encounter of a triple and a binary, or an encounter of a
binary with an interacting tangle of stars in which other multiples
have already temporarily been captured.  Following such complex
interactions poses severe challenges to the development of software
needed to choreograph these many-body dances.  Let us look at one such
system, Starlab, the most advanced in design for simulating dense
stellar systems.

\section{A Software Environment: Starlab}

Starlab is a software package for simulating the evolution of dense
stellar systems and analyzing the resultant data. It is a collection
of loosely coupled programs (`tools') linked at the level of the UNIX
operating system.  The tools share a common data structure and can be
combined in arbitrarily complex ways to study the dynamics of star
clusters and galactic nuclei.  Although most tools are written in C++,
Fortran programs and subroutines can be easily included, as well as
programs written in other languages.  The package is publicly
available through our web site $<$http://www.manybody.org$>$, where
links can be found to Starlab, as well as to other packages, such as
the Nemo environment (for collisionless stellar dynamics).  For a
description of the main Starlab components, see Portegies Zwart {\it et al.}
(2001).

\noindent
Starlab features the following basic modules:
\begin{itemize}
\item
Three- and four-body automated scattering packages, constructed
around a time-symmetrized Hermite integration scheme.
\item
A collection of initialization and analysis routines for use with
general N-body systems.
\item
A general Kepler package for manipulation of two-body orbits.
\item
N-body integrators incorporating both 2nd-order leapfrog and
4th-order Hermite integration algorithms.
\item
{\it Kira}, a general N-body integrator that automatically applies
local coordinate transformations in a recursive way, allowing uniform
treatment of hierarchical systems of arbitrary complexity within a
general N-body framework.
\item
{\it SeBa}, a stellar evolution and binary evolution package that
follows the evolution of any star or binary from arbitrary initial
conditions, or from later starting conditions following a dynamical
interaction.
\end{itemize}

A novel aspect of Starlab is its flexible external data representation,
which guarantees that tools can be combined in arbitrary ways, without
loss of data or internally-generated comments. Thus, two tools
connected by UNIX pipes may operate on different portions of the same
data set, even though neither understands the data structures, or even
the physical variables, used by the other.  Unknown data are simply
passed through unchanged to the next tool in the chain.

This feature is especially important for simulations that include the
use of both Kira and SeBa.  Although the two programs are linked at
compile time, before being run, we have gone to great length to
guarantee maximal decoupling between the structures of the two programs.
Conceptually we have the following picture:

\setlength{\unitlength}{2.4cm}

\bigskip
\begin{picture}(5,0.1)\thicklines
\put(0.5,0){\large Stellar Dynamics}
\put(3.2,0){\large Stellar Evolution}
\put(2.5,0.05){\vector(-1,0){0.4}}
\put(2.5,0.05){\vector(1,0){0.4}}
\end{picture}
\bigskip

\noindent
where the data passing currently takes place via an exchange
through a well-defined and minimal interface between the C++ objects
that represent the dynamical part of a star and the stellar evolution
aspects of that same star.

We are currently planning the next great leap forwards, which will
entail three significant improvements.  The first is to replace stellar
evolution recipes by actual stellar evolution codes.  The second is
to replace code linking by the more flexible mechanism of message passing.
The third is to introduce some form of stellar hydrodynamics code, to model
the details of stellar collisions.  Conceptually, our goal is to model
a dense stellar system as follows:

\bigskip
\bigskip
\begin{picture}(5,1)\thicklines
\put(0.5,0.95){\large Stellar Dynamics}
\put(3.2,0.95){\large Stellar Evolution}
\put(1.9,0){\large Hydrodynamics}
\put(2.5,1){\vector(-1,0){0.4}}
\put(2.5,1){\vector(1,0){0.4}}
\put(1.7,0.5){\vector(4,-3){0.4}}
\put(1.7,0.5){\vector(-4,3){0.4}}
\put(3.3,0.5){\vector(4,3){0.4}}
\put(3.3,0.5){\vector(-4,-3){0.4}}
\end{picture}
\bigskip

Let us look at each of these three aspects -- stellar evolution,
hydrodynamics, and code communication -- in turn. 

\section{Truly Interacting Stars}

It is a well kept secret that stellar evolution codes cannot follow
the full lifetime evolution of a star from beginning to end without
human intervention.  Modeling the main sequence evolution presents no
difficulties, but rapid changes occurring in later stages of stellar
evolution will either lead to a crash of the code or to such a slow
crawl that the run cannot be completed in a reasonable time.  This
does not pose a serious problem when we want to make a detailed study
of a particular star.  In that case we may not be interested in the
most unstable phases of its evolution; or if we are, we can help the
code along by some suitable choice of parameter changes when necessary.

This state of affairs does pose a serious problem, however, when we
want to use a stellar evolution code as a module in a larger system of
a hundred thousand stars or more.  In that case it would be totally
impractical to tweak the late stages of evolution of individual stars
by hand.  The obvious solution is to simplify stellar evolution codes,
letting them skip over some of the most numerically sensitive episodes
in their evolution.  While this introduces errors, it is surely better
than having no information at all about the late-time behavior of the
stars.  And as long as we can keep track of the approximations
involved, we can try to decrease the errors in due time, when faster
future computers allow us to spend more CPU time on an individual
star.

If we only had to deal with single stars, one could argue that it
would be better to use the best codes available, together with a very
patient lab assistant who coaxes the code through the difficult stages
in its evolution, time and again, for a wide choice of parameters,
such as initial mass and initial chemical composition.  After producing
a multidimensional collection of stellar evolution tracks, the values
can be stored in tables, or complicated parameter fits can be made.
For each star in a star cluster simulation, the physical parameters
describing the stellar structure and composition of a star can then be
obtained automatically from the tables or fits.

While this may well be the best approach for dealing with single stars,
it will not work for interacting double stars.  As soon as one of the
stars transfers mass to the other star, the evolution of both stars
will begin to diverge from the standard tracks obtained for single stars.
What will happen next depends on several parameters: the mass of each
of the stars, the initial separation between the stars, the chemical
composition of each of the stars.  Even if we assume both stars to
have the same composition labeled only by metallicity, we suddenly
find ourselves with a four-parameter family of pairs of evolution tracks
that will need to be tabulated or fitted, in addition to those of the
single stars.

And as if that would not be sufficiently impractical, we also have to
deal with blue stragglers, stars that have been formed through a
merger of two or sometimes even more than two stars.  And then there
are triple stars, where the outer star can perturb the orbit of the
inner two stars, thereby modifying the rate of mass transfer considerably.
It is clear that the use of canned stellar evolution tracks is not going 
to help us here.  There seems to be no other choice than to enlist the
help of a real stellar evolution code, albeit a simplified one.
Such a stellar evolution module should be able to produce results that
are at least somewhat reasonable without any need for human
hand-holding, and in a rather short amount of computer time.

For example, when we follow $3\times10^5$ stars, of which 10\% are
part of a primordial binary, there is a potential need to model
$3\times10^4$ stars separately.  However, many of those binaries will
be wide enough to avoid contact, while others may merge quickly and
therefore do need much CPU time.  As a round estimate, perhaps 1\% of
the stars will need to be followed in great detail, during the
simulation of the full history of the star cluster.  If we have one or
more fast front end computers connected to a cluster of GRAPE-6
boards, the total star cluster simulation might be completed in a few
months.  Dividing this total CPU time of $10^7$ seconds over the
$3\times10^3$ problem stars, we are left with up to an hour per star
(even more if we have several CPUs working together on the
simulation).  This is more than enough for a simple modeling job, and it
is clear that CPU time is not the bottle neck.  The only real problem
lies in the software: there is an urgent need for simple but robust
stellar evolution codes.

It may even be feasible to model all stars, single or double, by
direct calculation.  This would take away the need for the additional
complexity in the form of tables and fitting functions.  Depending on
the number of CPUs in the front end, such a brute-force approach would
leave us with a minute or several minutes per star.  For single stars,
this may already suffice; and if not now, then in the near future,
with the next generation of faster computers.

\section{Hydrodynamics}

When two stars in a binary exchange mass in a stable way, on a thermal
time scale (with the donor star typically the less massive of the two),
it is relatively straightforward to couple two stellar evolution
programs together.  At the moment of first contact, we already know
the precise structure of each star, and if we continue to use a
simple spherical approximation for the shape of each star, we only
have to change the boundary conditions.  Mass shed from the donor star
will be taken from its outer layers, and added to the outer layers of
the receiving star.  Dynamical equilibrium can be easily enforced
right away, and the two stars will then take care of the restoration
of thermal equilibrium automatically, in the course of their
subsequent evolution.

Things get quite a bit more complex when two stars collide.  This can
occur during an unstable phase of mass transfer, for example when the
heaviest star transfers matter to the lighter star.  Conservation of
angular momentum then tends to decrease the distance between the
stars, leading in most cases to further mass transfer, resulting in a
spiral-in of the two star cores in their common envelope.  Other
scenarios leading to stellar collisions include random encounters
between single stars in the dense cores of stellar systems, as well as
local encounters within interacting systems of three or more stars.

The simplest example of the latter occurs when a binary and a single
star meet each other.  While there is already a chance for a collision
between two or perhaps all three stars during the initial encounter,
the collision probability is significantly enhanced when the three
stars are temporarily captured in what is called a resonance
scattering event ({\it cf.} Heggie \& Hut 2002).  If the stars would
truly be point masses, such an unstable three-body system would
ultimately fall apart again into a single star and a double star, but
the stars may conduct tens or hundreds of passes through the system
before finally escaping.  Even in relatively wide configurations, the
cumulative chance for collisions between at least two of the stars
during this protracted dance can be considerable.

In most star clusters a significant fraction of the stars are born in
binaries.  The probability for two of these primordial binaries to
meet each other is relatively high, given the fact that binaries tend to
sink to the core of a star cluster on a two-body relaxation time
scale, as a result of mass segregation.  When four stars meet each
other to form a temporarily bound subsystem, the chance for collisions
is even higher than in the three-body case, given that there are six
pair-wise permutations for which collisions can take place.

The first case mentioned, that of common-envelope evolution, is the
hardest one to model.  We are dealing with a three-dimensional problem
in hydrodynamics, where the initial stages of slow spiral-in can cover
a very large number of orbital periods.  In contrast, collisions in
the other scenarios are almost instantaneous events.  The effects of a
collision between two stars make themselves felt within the interiors
of the stars on a sound crossing time, of order an hour for a main
sequence star.  Even if a single star passes through a red giant, the
collision between the star and the core of the giant will take place
on a time scale of hours; and if the star misses the core, the chance
is high that no merger will take place at all.

Collisions can be modeled with modest accuracy by using a simple
hydrodynamics code, preferably an SPH code.  Here SPH stands for
Smooth Particle Hydrodynamics, where each particle carries not only a
mass, position and velocity, as in stellar dynamics, but also an
entropy.  With each particle thus representing a fluid element, it is
easy to model systems with extreme density contrasts, especially if
different particles are allowed to have different masses.  With a
dynamic particle splitting and merging scheme, it then becomes
possible to model both the low-density exteriors as well as the
high-density cores of stars involved in a collision.

The central idea is to represent each star by a single mass point
together with a stellar evolution module (either a canned stellar
evolution track or an simple real-time stellar evolution code), during
most of the lifetime of a star.  Only just before a strong encounter
will the single mass point be replaced by a whole cloud of SPH
particles.  The internal structure of the SPH object, such as density,
temperature and chemical composition as a function of radius, can be
determined from the information carried in the stellar evolution
module.  After the strong encounter is finished, the SPH objects
should be given some extra time to settle down on a dynamical time
scale.  Finally, each object can be collapsed back into a point, to
take part once again in the overall dynamical $N$-body dance of the
star cluster as a whole.

A major complication in this last stage of transformation is the
question of how to update the stellar evolution module of each SPH
object.  The radial distribution of temperature, density and chemical
composition can be read off with limited accuracy from the cloud of
SPH particles.  The challenge is then to construct a new stellar
structure model for the object, as a new starting model for the
subsequent stellar evolution calculations.  Since the stellar
structure at this point is likely to be far off from thermal
equilibrium, there may not be any standard model available to use as a
guiding tool.  Again, we must insist on the stellar evolution code to
be smart enough to find a new starting model anyway, without any human
guidance.

\section{Message Passing}

Our first goal is to have primitive working versions in place of an
hydrodynamics code and a stellar evolution code, together with an
implementation of the rather complex administrative rules that guide
their interaction with each other and with the leading stellar dynamics
code.  While this already poses quite a few challenges, it is only
part of what is required for a full simulation of a star cluster.  To
make everything work, we have to find ways to couple everything together.

Traditionally, this is done by linking the various modules together at
compile time.  Each module then acts pretty much as a subroutine in a
Fortran program.  And while each module can be compiled and tested
separately, in the end they together form one massive program where in
principle everything is coupled.  For an intermediate-size software
project, this may not be such a bad thing.  A good programming style
can go a long way to ensure some form of data abstraction and
separation.  For example, avoiding the use of global variables (or
common blocks when programming in Fortran) would be a good start.

However, the larger the total program becomes, the greater are the
dangers of unintended interactions between its parts.  This will make
debugging increasingly difficult.  Since different people will write
and maintain different modules, debugging will probably require
interactive collaboration between those individuals, further
complicating overall code maintenance.  Clearly, by the time we have
complex working codes for stellar dynamics, stellar evolution, and
stellar hydrodynamics, we have to look for better ways to integrate
everything.

In theoretical astrophysics, so far there has been little need to go
beyond the model of working with a handful of program files, linked
together at compile time.  But in various areas of software
development, starting a couple decades ago, other solutions have been
explored in order to deal with increasingly complex configurations of
interacting programs.  An example is the X window system developed at
M.I.T. in the early to mid eighties.  Since X is open source, it has
been extended widely every since; see $<$http://www.X.org/about\_x.htm$>$
for a brief introduction.  The core ideas in X are to run the
different modules asynchronously, without a single core program
telling each module what to do and when.

This lack of centralized control allows much greater freedom for the
construction and maintenance of individual modules.  Each module can be
regarded as a black box with a well-defined protocol that allows for
exchange of information between the boxes.  Communication takes place
through message passing, in a client-server model.  One module can
send a message to another module, which then leads to the appropriate
action as soon as possible given the other demands on that module.
As with any form of parallelism, an appropriate tuning of priority
levels (hopefully) leads to a reasonable minimization of overall
waiting times, depending on the ramifications of each request.

We are currently contemplating a somewhat similar implementation for a
master framework that will allow stellar dynamics, evolution, and
hydro codes to talk to each other.  Such a framework, once in place,
will make it trivial to mix and match modules written in different
styles and in different computer languages.  All that is needed for a
new module to be installed, or for a current module to be replaced, is
to agree upon the interface protocol, including the set of data that
can be exchanged and their precise formatting.  We hope to start this
development, from X windows to `windows on the Universe' in the near
future.

\section{An Extended Software Environment}

Currently, Starlab does not have a fully working hydrodynamics module,
although we have experimented with some toy models as a proof of
concept for translating $N$-body particles into polytropic star models
realized in SPH, and {\it vice versa}.  Also, Starlab has not yet
incorporated real stellar evolution codes.  So far, all effects related
to stellar structure and evolution have been handled by the SeBa
package mentioned in section 2, which relies on formulas fitted to
previously computed stellar evolution tracks (such as given by Hurley
{\it et al.} 2000).

Even with those restrictions still in place, we have recently
encountered the need for additional software improvements, mainly as a
result of a vast increase in data throughput.  The source of this good
news is the fact that a large number of GRAPE-6 boards have come
on-line, in the form of the 48-Teraflops cluster in Tokyo as well as
smaller clusters elsewhere in the world.  This has made it possible
for us to model small globular clusters on a star-by-star basis for
the duration of a Hubble time.  However, the Terabytes of data
produced by those runs have posed new challenges in the areas of
flexible visualization, efficient archiving, and the exchange of data
with other groups.  Let us briefly discuss each of these three
challenges (see figure 1 for an overview).

\subsection{Visualization}

\begin{figure}[t]
\begin{center}
\leavevmode
\epsfxsize = 12cm
\epsffile{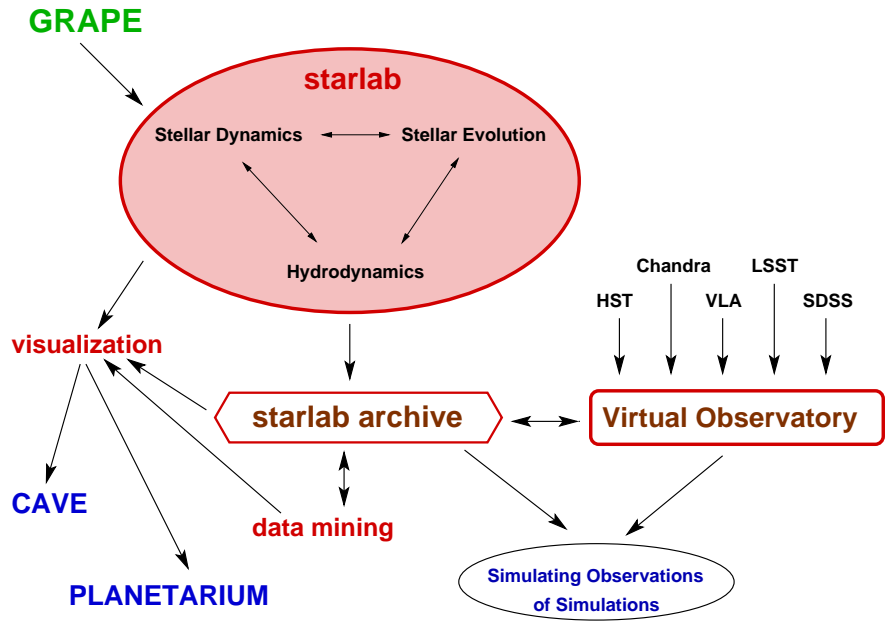}
\caption{A software environment for simulations of dense stellar
systems.  Centered around Starlab, it contains connections with a
library of GRAPE routines, with various visualization tools, and with
an archive to store the most interesting runs.  In addition, it will
maintain links to virtual observatory initiatives.}
\label{fig:vo}
\end{center}
\end{figure}

We have started to explore the possibility of using planetarium domes
to create a fully immersive interactive virtual reality environment to
display the 4-D space-time data from a complete globular cluster
simulation history.  Using software developed at NCSA for their CAVE
({\it cf.} $<$http://evlweb.eecs.uic.edu/pape/CAVE$>$), and later
adapted to the dome in the Hayden Planetarium at the American Museum
of Natural History, we have been able to `fly through' our star
clusters in space and time.  In contrast to a computer screen, which
forces the user to zoom in and out repeatedly in order to get a
complete picture of what is going on, a dome allows the user to retain
the global view while at the same time focusing on individual
interaction regions.  The extra resolution afforded by a set of seven
projectors (two covering the `polar cap', and five covering the
azimuthal sweep along the `horizon') makes this simultaneous
local-global investigation possible.  See Teuben (2002) and Teuben
{\it et al.} (2001, 2002) for more details.

\subsection{Archives}

With the ability to visualize the complete history of a small globular
cluster on a star-by-star and orbit-by-orbit basis, it would be a
shame not to store the appropriate data set from, say, a month-long run on
a GRAPE-6 cluster.  Fortunately, not every position and velocity value
has to be stored for each star at each time step.  Experimentation has
shown that storing only a few percent of those data suffices to
reconstruct the orbits through higher-order interpolation, to a degree
that is virtually indiscernible from the real thing, to the human eye.
Even so, this leaves us with the need to archive Terabytes of data for
our largest runs, in such a form that it will be relatively easy to
extract information sliced in space as well as in time.  In some cases,
we may want to display snapshots of the whole system.  In other cases,
we may be interested on what happened to only a handful of interacting
stars, but over a much longer period of time.  We are currently
experimenting with different ways to solve these conflicting demands
efficiently.

\subsection{Virtual Observatories}

The challenge of dealing with enormous data sets has hit observational
astronomy long before it has become a problem for theoretical
simulations.  In various countries new initiatives have started to
address these problems through the construction of what are called
virtual observatories.  These will provide portals on the internet
that allow the user to become a virtual observer, `observing' data
available in archives from major observatories, ground-based as well
as space-based.  In addition, the user will be able to correlate
information from and submit queries to a diverse range of sources.
All this is still in its initial stages, but the results so far are
already promising.  See the proceedings of the recent conference {\it
Virtual Observatories of the Future} (Brunner{\it et al.} 2001), or
the NVO website $<$http://nvosdt.org$>$ for pointers to the
U.S. initiative as well as to parallel efforts in other countries.

\section{Summary}

We are now at the verge of simulating the full history of a globular
cluster on a star-by-star basis, following each star dynamically
while also modeling their stellar evolution.  Initially we will use
100,000 stars, and we hope to extend this number to 300,000 or more;
by the end of the decade, with Petaflops computers such as the GRAPE-8,
we may reach our holy grail of accurately modeling a million stars for
10 billion years, while resolving local interactions involving binaries
and more complex systems on time scales of hours and days.

We expect to deliver two rather different types of results, one
oriented to observations, one to theory.  On the observational side,
we would like to give visitors the option to look at our simulated
clusters at any time and distance of their choice, through whatever
instrument they like (say, in a given wave length band, using the
Hubble Space Telescope, at a distance of 5 kpc when the cluster was
8 billion years old; or using Chandra, the VLA, etc.).

On the theoretical side, we can use our largest runs to provide access
to other theorists who have more specific questions about particular
objects.  For example, if we are interested in finding out how binary
pulsars are formed, through what channels and with what type of
eccentricity distributions, we have no choice but to run a full
simulation of a star cluster, and then to pick out the binary pulsar
related events.  However, this same simulation will allow others to
analyse in detail what happened with the CVs in the system, the blue
stragglers, as well as many other variable stars and binaries,
triples, etc.  So we will invite not only `guest observers', but also
`guest theorists' who can pick their favorite objects from our
simulation and compare our space-time trajectories and interactions of
those objects with their own analytical and/or numerical attempts to
model them in isolation.  Their modeling may be more detailed but
we provide a more realistic context, thus providing complementary ways
to shed light on the objects of interest.

We currently have three essential ingredients in place to do these
jobs: access to a super-fast computer, the GRAPE-6 running at 48
Teraflops in Tokyo; the software to run our simulations in the form of
the Starlab environment; and an opportunity to visualize our results
using the Hayden Planetarium dome at the American Museum of Natural
History in New York City.  What is needed now is further software
development, both within and around Starlab.  Within Starlab we plan
to introduce stellar evolution codes as well as hydrodynamics (SPH)
codes, together with a message passing framework to connect both type
of codes with each other and with the Kira code that governs the
stellar dynamics.  Around Starlab we plan to improve our use of
partiview, the open source version of the Virtual Director software
used in the Hayden dome; to create Starlab archives; to develop
software for guest observers; and to establish links to virtual
observatories.

\bigskip
I acknowledge support from the Alfred P. Sloan foundation, through a
grant for research at the Hayden Planetarium of the American Museum of
Natural History in New York.

\def\pasj{PASJ}
\def\mn{MNRAS}
\def\MN{MNRAS}
\def\mnras{MNRAS}
\def\apj{ApJ}
\newcommand{\etalchar}[1]{$^{#1}$}
\newcommand{\noopsort}[1]{} \newcommand{\printfirst}[2]{#1}
  \newcommand{\singleletter}[1]{#1} \newcommand{\switchargs}[2]{#2#1}

\end{document}